\documentclass[twoside,bezier,12pt, reqno]{amsart}
\usepackage{amssymb,latexsym,epsfig,mathrsfs,graphpap,graphics,amsmath,amssymb}
\usepackage{amstext}
\usepackage{amsthm}
\usepackage[utf8]{inputenc}
\usepackage[english]{babel}
\usepackage{helvet}
\usepackage{courier}

\newtheorem{theorem}{Theorem}[section]
\newtheorem{lemma}{Lemma}[section]

\theoremstyle{Definition}
\newtheorem{definition}{Definition}[section]

\theoremstyle{remark}
\newtheorem{remark}[theorem]{Remark}
\numberwithin{equation}{section}

\begin{document}

\begin{flushleft}
 {\bf\Large {Quaternion  Offset Linear Canonical  Transform in One-dimensional Setting}}

\parindent=0mm \vspace{.2in}

{\bf{M. Younus Bhat$^{1},$ and Aamir H. Dar$^{2}$ }}
\end{flushleft}

{{\it $^{1}$ Department of  Mathematical Sciences,  Islamic University of Science and Technology Awantipora, Pulwama, Jammu and Kashmir 192122, India.E-mail: $\text{ gyounusg@gmail.com}$}}

{{\it $^{2}$ Department of  Mathematical Sciences,  Islamic University of Science and Technology Awantipora, Pulwama, Jammu and Kashmir 192122, India.E-mail: $\text{ahdkul740@gmail.com}$}}

\begin{quotation}
\noindent
{\footnotesize {\sc Abstract.}  In this paper, we introduce quaternion offset linear canonical transform of integrable and square integrable functions. Moreover, we show that the proposed transform satisfies all the respective properties like inversion formula, linearity, Moyal's formula , product theorem and the convolution theorem.  \\


\noindent
\textit{2000 Mathematics subject classification: } }
\end{quotation}

\section{\bf Introduction}
\noindent
The classical Integral transform has been generalized to the six-parameter $(A, B, C, D, p, q)$ transform called the offset linear canonical transform (OLCT). 
For a matrix parameter $\Lambda =\left[\begin{array}{cccc}A & B &| & p\\C & D &| & q \\\end{array}\right] $,  the OLCT of any signal $f$ is defined as
\begin{equation}\label{eqn 1}
\mathcal O_\Lambda[f](w)=\int f(t)\mathcal K_{\Lambda}(t,w)dt,
\end{equation}
where $\mathcal K_{\Lambda}(t,w)$ denotes the   kernel of the OLCT and is  given by
\begin{equation}
\mathcal K_{\Lambda}(t,w) =\frac{1}{\sqrt{i2\pi B}}\exp\left\{\frac{i}{2B}\left(At^2-2t(w-p)-2w(Dp-Bq)+D(w^2+p^2)\right)\right\}
\end{equation}
with $AD-BC=1.$

It is here worth to mention that if  $ B=0$  then the OLCT  defined by  (\ref{eqn 1}) is simply a time
scaled version off multiplied by a linear chirp. Hence, without loss of generality, in this paper  we
assume $B\ne0$.

Looking at the core of OLCT, one can derive it as a time-shifted and frequency-modulated child of the parent linear canonical transform (LCT) \cite{xulct, 16l}. On the application side OLCT is similar to LCT but due to its two extra parameters $p$ and $q$, it is more general and flexible than parental LCT. Hence it has gained more popularity in optics, signal and image processing. For more details, we refer to \cite{15l, 14, xiang}.

In the prospect of signal processing, one can consider that any
signal processing tool converts the time-domain signals into frequency-domain . Further in signal processing, convolution of two
functions \cite{17l,18l,ownwvd} is a most useful tool in constructing a filter for denoising the given noisy signals(see\cite{19l}).\\

In past decades, hypercomplex algebra has become a leading area of research with its applications in color image processing, image filtering, watermarking, edge detection and pattern recognition(see \cite{1,2,3,10,4,5,6,7}). The Cayley-Dickson algebra of order four is labeled as quaternions which has wide applications in  optical and signal processing. The extension of Fourier transform in quaternion algebra is known as quaternion Fourier transform(QFT) \cite{8} which is said to be the substitute of the commonly used two-dimensional Complex Fourier Transform (CFT). The QFT has wide range  of applications see\cite{11,12}. 

The quaternionic offset linear canonical transform (QOLCT) can be defined as a generalization of the quaternionic linear canonical transform (QLCT) and has been studied in \cite{bah2}. Here the authors derive the relationship between the QOLCT and the quaternion Fourier transform (QFT). Moreover, they proved the Plancherel formula, and some properties related to the QOLCT. For more details we refer to \cite{bia,13,ownqgolct,ownd}.

But to the best of our knowledge,  theory about one-dimensional quaternion OLCT(1D-QOLCT)  is still in its infancy.  Therefore it is worthwhile to study the theory of 1D-QOLCT which can be productive for signal processing theory and applications.
    In this paper, our main objectives are to introduce the novel integral transform called the one-dimensional quaternion offset linear canonical  transform(1D-QOLCT) and study its properties, such as inversion formula, linearity,
Moyal's formula, convolution theorem and product theorem 1D-QOLCT.\\
    This paper is organized as follows: In Section 2, we summarize  the general definitions and basic properties of quaternions.  In Section 3, we introduce 1D-QOLCT and obtain various properties linearity, Moyal's formula, convolution and product theorem of the proposed transform.

\section{\bf Preliminaries}
\subsection{Quaternions}\ \\
Let $\mathbb R$ and $\mathbb C$ be the usual set of real numbers and set of complex numbers, respectively. The division ring of quaternions in the honor of Hamilton, is denoted by $\mathbb H$ and is defined as
 \begin{align*}
 \mathbb H&=\{h_0+e_1h_1+e_2h_2+e_3h_3:h_0,h_1,h_2,h_3\in\mathbb R \}\\
 &=\{z_1+e_2z_2:z_1,z_2\in\mathbb C\}\quad(Cayley Dickson form)
 \end{align*}
 where $e_1,e_2,e_3$ satisfy Hamilton’s multiplication rule\\
$$ e_1e_2=-e_2e_1=e_3,\quad e_2e_3=-e_3e_2=e_1,\quad e_3e_1=-e_1e_3=e_2$$
and\\
$$e^2_1=e^2_2=e^2_3=1$$

\parindent=0mm \vspace{.2in}

Every member of  $\mathbb H$ is known as quaternion. In quaternion algebra addition, multiplication, conjugate
and absolute value of quaternions are defined by \\
$$(a_1+e_2a_2)+(b_1+e_2b_2)=(a_1+b_1)+e_2(a_2+b_2),$$
$$(a_1+e_2a_2)(b_1+e_2b_2)=(a_1b_1-\overline a_2b_2)+e_2(a_2b_1+\overline a_1 b_2), $$
$$(a_1+e_2a_2)^c=\overline a_1-e_2a_2,$$
$$|a_1+e_2a_2|=\sqrt{|a_1|^2+|a_2|^2},$$\\
here $\overline a_k$ is the complex conjugate of $a_k$ and $|a_k|$ is the modulus of the
complex number $a_k, k = 1, 2.$
 For all $a=a_1+e_2a_2,\quad b=b_1+e_2b_2 \in \mathbb H,$ the following properties of conjugate and modulus and multiplicative inverse
are well known.\\
$$(a^c)^c=a,\quad (a+b)^c=a^c+b^c,\quad(ab)^c=b^ca^c,$$ $$\quad|a|^2=aa^c=|a_1|^2+|a_2|^2,\quad|ab|=|a||b|,$$
$$a^{-1}=\frac{\overline a}{|a|^2}.$$\\

We denote   $L^p(\mathbb R,\mathbb H),$ the Banach space of all quaternion-valued functions $f$
satisfying $$\|f\|_p=\left(\int|f_1(t)|^p+|f_2(t)|^pdt\right)^{1/p}<\infty,\quad p=1,2.$$ And  on $L^2(\mathbb R,\mathbb H)$ the inner product $\langle f,g\rangle=\int f(t)[g(t)]^cdt,$ where integral of a quaternion valued function is defined by $\int(f_1+e_2f_2)(t)dt=\int f_1(x)dt+e_2\int f_2(x)dt,$ whenever the integral exists.

\section{\bf Quaternion one-dimensional  offset linear canonical  transform }

 In this section we will introduce  the definition of quaternion one-dimensional offset linear canonical transform(1D-QOLCT) by using \cite{1D-QFT,1D-QFrFT,1D-QLCT}. Prior to that we note $e_1,e_2$  and $e_3$ ((or equivalently $i,j,k$) denote the three imaginary units in the quaternion algebra.

\begin{definition}\label{def 1D-QOLCT} The 1D-QOLCT of any signal $f\in L^1(\mathbb R,\mathbb H)$ with respect a matrix parameter $\Lambda=(A,B,C,D,p,q)$ is defined by
\begin{equation}\label{eqn 1D-QOLCT}
\mathbb Q^{\mathbb H}_{\Lambda}[f(t)](w)=\int f(t)\mathcal K^{ e_2}_{\Lambda}(t,w)d{ t}
\end{equation}
where
\begin{equation}
\mathcal K_{\Lambda}(t,w) =\frac{1}{\sqrt{i2\pi B}}\exp\left\{\frac{i}{2B}\left(At^2-2t(w-p)-2w(Dp-Bq)+D(w^2+p^2)\right)\right\}
\end{equation}
With $AD-BC=1.$
Now we can find that if $f(t)$ is real-valued signal  in (\ref{eqn 1D-QOLCT}), then we can interchange the kernel in Definition \ref{def 1D-QOLCT}.
\end{definition}

By appropriately choosing parameters in $\Lambda=(A,B,C,D,p,q)$  the 1D-QOLCT(\ref{eqn 1D-QOLCT})
gives birth to the following existing time-frequency transforms:
\begin{itemize}
\item For $\Lambda=(0,1,-1,0,0,0)$, the 1D-QOLCT (\ref{def 1D-QOLCT}) boils down to the quaternion one-dimensional  Fourier Transform\cite{1D-QFT}

\item  For $\Lambda=(A,B,C,D,0,0)$  the 1D-QOLCT(\ref{def 1D-QOLCT}) reduces to the Quaternion one-dimensional Linear Canonical Transform\cite{1D-QLCT}.\\
\item  For $\Lambda=(\cos\theta,\sin\theta,-\sin\theta,\cos\theta,0,0)$  the 1D-QOLCT (\ref{def 1D-QOLCT}) reduces to the Quaternion one-dimensional fractional Fourier Transform\cite{1D-QFrFT}.
\end{itemize}

\begin{definition}[Inversion]\label{def inverse}
The inverse of a 1D-QOLCT with parameter $\Lambda =\left[\begin{array}{cccc}A & B &| & p\\C & D &| & q \\\end{array}\right] $ is given by a 1D-QOLCT with parameter  $\Lambda^{-1} =\left[\begin{array}{cccc}D& -B &| &Bq-Dp\\-C & A &| &Cp-Aq \\\end{array}\right] $  as
\begin{equation}\label{eqn inverse}
f(t)=\{\mathcal O^{\mathbb H}_{\Lambda}\}^{-1}[\mathcal O^{\mathbb H}_{\Lambda}[f]](t)
=\int\mathcal O^{\mathbb H}_{\Lambda}[f](w){\mathcal K^{ e_2}_{\Lambda^{-1}}(w,t)}dw\\
\end{equation}
where ${\mathcal K^{ e_2}_{\Lambda^{-1}}(w,t)}={\mathcal K^{ -e_2}_{\Lambda}(t,w)}=\overline{\mathcal K^{ e_2}_{\Lambda}(t,w)}$
\end{definition}

\begin{definition}\label{def2 1D-QOLCT}
Let $f=f_1+e_2f_2$ be a quaternion valued signal in $L^1(\mathbb R,\mathbb H),$ then the quaternion quadratic-phase Fourier transform is defined as
\begin{equation}\label{eqn2 1D-QOLCT }
\mathcal O^{\mathbb H}_{\Lambda}[f(t)]( w)=\mathcal O^{\mathbb H}_{\Lambda}[f_1(t)]( w)+e_2\mathcal O^{\mathbb H}_{\Lambda}[f_2(t)]( w).
\end{equation}
By above definition, it is consistent with the offset linear canonical transform on $L^1(\mathbb R,\mathbb C).$ Now it is clear from the definition of quaternion offset linear canonical transform  and the properties of offset linear canonical transform on $L^1(\mathbb R,\mathbb H),$ that $\mathcal O^{\mathbb H}_{\Lambda}\left(\mathcal O^{\mathbb H}_{\Gamma}[f]\right)=\mathcal O^{\mathbb H}_{\Lambda\Gamma}[f]$  and $\{\mathcal O^{\mathbb H}_{\Lambda}[f]\}^{-1}=\mathcal O^{\mathbb H}_{\Lambda^{-1}}[f]$ for every signal $f\in L^1(\mathbb R,\mathbb H).$
\end{definition}

\begin{theorem}\label{linear} The quaternion quadratic-phase Fourier transform $\mathcal O^{\mathbb H}_{\Lambda}$ is $\mathbb H-$linear on $L^1(\mathbb R,\mathbb H).$
\begin{proof}
Let us consider two quaternion signals $f=f_1+e_2f_2$ and $g=g_1+e_2g_2$ in $L^1(\mathbb R,\mathbb H),$ now by the linearity of  $\mathcal O^{\mathbb H}_{\Lambda}$ on $L^1(\mathbb R,\mathbb C),$  we obtain
\begin{align*}
\mathcal O^{\mathbb H}_{\Lambda}[f+g]&=\mathcal O^{\mathbb H}_{\Lambda}[(f_1+e_2f_2)+(g_1+e_2g_2)]\\
&=\mathcal O^{\mathbb H}_{\Lambda}[(f_1+g_1)+e_2(f_2+g_2)]\\
&=\mathcal O^{\mathbb H}_{\Lambda}[f_1]+\mathcal O^{\mathbb H}_{\Lambda}[g_1]+e_2\left(\mathcal O^{\mathbb H}_{\Lambda}[f_2]+\mathcal O^{\mathbb H}_{\Lambda}[g_2]\right)\\
&=\left(\mathcal O^{\mathbb H}_{\Lambda}[f_1]+e_2\mathcal O^{\mathbb H}_{\Lambda}[f_2]\right)+\left(\mathcal O^{\mathbb H}_{\Lambda}[g_1]+e_2\mathcal O^{\mathbb H}_{\Lambda}[g_2]\right)\\
&=\mathcal O^{\mathbb H}_{\Lambda}[f]+\mathcal O^{\mathbb H}_{\Lambda}[g].
\end{align*}
Now to prove $\mathbb H-$linearity, we let $q=q_1+e_2q_2 \in\mathbb H$ and $f=f_1+e_2f_2\in L^1(\mathbb R,\mathbb H)$ be arbitrary,then we have
\begin{align*}
\mathcal O^{\mathbb H}_{\Lambda}[e_2f]&=\mathcal O^{\mathbb H}_{\Lambda}[e_2(f_1+e_2f_2)]\\
&=\mathcal O^{\mathbb H}_{\Lambda}[e_2f_1-f_2]\\
&=e_2\mathcal O^{\mathbb H}_{\Lambda}[f_1]-\mathcal O^{\mathbb H}_{\Lambda}[f_2]\\
&=e_2\left(\mathcal O^{\mathbb H}_{\Lambda}[f_1]+e_2\mathcal O^{\mathbb H}_{\Lambda}[f_2]\right)\\
&=e_2\mathcal O^{\mathbb H}_{\Lambda}[f].
\end{align*}
Therefore,
\begin{align*}
\mathcal O^{\mathbb H}_{\Lambda}[qf]&=\mathcal O^{\mathbb H}_{\Lambda}[q_1f]+\mathcal O^{\mathbb H}_{\Lambda}[e_2q_2f]\\
&=q_1\mathcal O^{\mathbb H}_{\Lambda}[f]+e_2q_2\mathcal O^{\mathbb H}_{\Lambda}[f]\\
&=(q_1+e_2q_2)\mathcal O^{\mathbb H}_{\Lambda}[f]\\
&=q\mathcal O^{\mathbb H}_{\Lambda}[f].\\
\end{align*}
Which completes proof.
\end{proof}
\end{theorem}

\begin{theorem}[Moyal's formula ]\label{moyal} Let $f,g\in L^1(\mathbb R,\mathbb H)\cap L^2(\mathbb R,\mathbb H)$  be two signals functions with $\mathcal O^{\mathbb H}_{\Lambda}[f]\in L^1(\mathbb R,\mathbb H)$, $\langle f,g \rangle=\langle\mathcal O^{\mathbb H}_{\Lambda}[f],\mathcal O^{\mathbb H}_{\Lambda}[g] \rangle$ .
\begin{proof}
For  $f,g\in L^1(\mathbb R,\mathbb H)\cap L^2(\mathbb R,\mathbb H)$ with $\mathcal O^{\mathbb H}_{\Lambda}[f]\in L^1(\mathbb R,\mathbb H)$,
\begin{align*}
\langle f,g \rangle &=\int f(t)[g(t)]^cdt\\\\
&=\int\int\mathcal O^{\mathbb H}_{\Lambda}[f](w){\mathcal K^{ e_2}_{\Lambda^{-1}}(w,t)}dw[g(t)]^cdt\quad(by \ref{eqn inverse})\\\\
&=\int\int\mathcal O^{\mathbb H}_{\Lambda}[f](w){\mathcal K^{ -e_2}_{\Lambda}(t,w)}[g(t)]^cdwdt\\\\
&=\int\mathcal O^{\mathbb H}_{\Lambda}[f](w)\left\{\int{\overline{\mathcal K^{ e_2}_{\Lambda}(t,w)}}[g(t)]^cdw\right\}dt\\\\
&=\int\mathcal O^{\mathbb H}_{\Lambda}[f](w)\left\{\int{g(t)\mathcal K^{e_2}_{\Lambda}(t,w)}dw\right\}^cdt\\\\
&=\int\mathcal O^{\mathbb H}_{\Lambda}[f](w)\{\mathcal O^{\mathbb H}_{\Lambda}[g](w)\}^cdt\\\\
&=\langle\mathcal O^{\mathbb H}_{\Lambda}[f],\mathcal O^{\mathbb H}_{\Lambda}[g] \rangle.
\end{align*}
Which completes proof.
\end{proof}
\end{theorem}

\begin{lemma}\label{lem conju}
For  $f\in L^p(\mathbb R ,\mathbb C ),p=1,2$; we have $\mathcal O^{\mathbb H}_{\Lambda}[\overline{f}](w)=\overline{\mathcal O^{\mathbb H}_{\Lambda^{-1}}[f](w)}.$
\begin{proof}
It follows  from definition \ref{def 1D-QOLCT} that
\begin{align*}
\mathcal O^{\mathbb H}_{\Lambda}[\overline{f}](w)&=\int\overline{f(t)}\mathcal K^{e_2}_\Lambda(t,w)dt\\\\
&=\int\overline{f(t)\overline{\mathcal K^{e_2}_\Lambda(t,w)}}dt\\\\
&=\int\overline{f(t){\mathcal K^{-e_2}_{\Lambda}(t,w)}}dt\\\\
&=\overline{\int{f(t){\mathcal K^{e_2}_{\Lambda^{-1}}(w,t)}}dt}\\\\
&=\overline{\mathcal O^{\mathbb H}_{\Lambda^{-1}}[f](w)}.
\end{align*}
Which completes the proof.
\end{proof}
\end{lemma}
\begin{remark}The Lemma \ref{lem conju} can also be written as  $\mathcal O^{\mathbb H}_{\Lambda^{-1}}[\overline{f}](w)=\overline{\mathcal O^{\mathbb H}_{\Lambda}[f](w)}.$
\end{remark}
\begin{definition}\label{def con}
For $f\in L^2(\mathbb R,\mathbb H)$ and $g\in L^1(\mathbb R,\mathbb H),$ define
\begin{equation}
(f\ast g) =(f_1\ast g_1 -\mathcal O^{\mathbb H}_{\Lambda^{-2}}[\overline f_2\ast g_2])+e_2(f_2\ast g_1+\mathcal O^{\mathbb H}_{\Lambda^{-2}}[\overline f_1\ast g_1]),
\end{equation}
where $ \ast$ is the proposed definition of convolution.
\end{definition}
\begin{lemma} \label{lemma con} Under the assumptions of definition \ref{def con}, we have
\begin{equation}
\mathcal O^{\mathbb H}_{\Lambda}[f\ast g(t)](u)=\mathcal O^{\mathbb H}_{\Lambda}[f](u)\mathcal O^{\mathbb H}_{\Lambda}[ g](u)\exp\left\{\frac{e_2}{2B}\left(2w(Dp-Bq)-Dw^2\right)\right\}
\end{equation}
\end{lemma}

\begin{theorem}[Convolution theorem]\label{thm con} Let $f,g$ be two given signal functions such that $f\in L^2(\mathbb R,\mathbb H)$ and $g\in L^1(\mathbb R,\mathbb H),$ , then  for all $w\in \mathbb R$ we have
\begin{equation}
\mathcal O^{\mathbb H}_{\Lambda}[f\ast g](w)=\mathcal O^{\mathbb H}_{\Lambda}[f](w)\mathcal O^{\mathbb H}_{\Lambda}[ g](w)\exp\left\{\frac{e_2}{2B}\left(2w(Dp-Bq)-Dw^2\right)\right\}
\end{equation}
\begin{proof}
By applying  Definition \ref{def con} and Lemma \ref{lemma con}, we have
\begin{align*}
\mathcal O^{\mathbb H}_{\Lambda}[f\ast g](w)&= \mathcal O^{\mathbb H}_{\Lambda}\left[(f_1\ast g_1 -\mathcal O^{\mathbb H}_{\Lambda^{-2}}[\overline f_2\ast g_2])\right](w)+e_2\mathcal O^{\mathbb H}_{\Lambda}\left[(f_2\ast g_1+\mathcal O^{\mathbb H}_{\Lambda^{-2}}[\overline f_1\ast g_1])\right](w)\\\\
&=\mathcal O^{\mathbb H}_{\Lambda}[f_1](w)\mathcal O^{\mathbb H}_{\Lambda}[g_1](w)\exp\left\{\frac{e_2}{2B}\left(2w(Dp-Bq)-Dw^2\right)\right\}-\mathcal O^{\mathbb H}_{\Lambda}\mathcal O^{\mathbb H}_{\Lambda^{-2}}[\overline f_2](w)\mathcal O^{\mathbb H}_{\Lambda}[g_2](w)\\\\
&\qquad\times\exp\left\{\frac{e_2}{2B}\left(2w(Dp-Bq)-Dw^2\right)\right\}\\\\
&+e_2\left\{\mathcal O^{\mathbb H}_{\Lambda}[f_2](w)\mathcal O^{\mathbb H}_{\Lambda}[g_1](w)\exp\left\{\frac{e_2}{2B}\left(2w(Dp-Bq)-Dw^2\right)\right\}\right.\\\\
&\qquad\left.+\mathcal O^{\mathbb H}_{\Lambda}\mathcal O^{\mathbb H}_{\Lambda^{-2}}[\overline f_1](w)\mathcal O^{\mathbb H}_{\Lambda}[g_1](w)\exp\left\{\frac{e_2}{2B}\left(2w(Dp-Bq)-Dw^2\right)\right\}\right\}\\\\
&=\left\{\big[\mathcal O^{\mathbb H}_{\Lambda}[f_1]\mathcal O^{\mathbb H}_{\Lambda}[g_1]- \mathcal O^{\mathbb H}_{\Lambda^{-2}}[\overline f_2]\mathcal O^{\mathbb H}_{\Lambda}[g_2]\big](w)\right.\\\\
&\qquad\left.e_2\big[\mathcal O^{\mathbb H}_{\Lambda}[f_2]\mathcal O^{\mathbb H}_{\Lambda}[g_1]- \mathcal O^{\mathbb H}_{\Lambda^{-1}}[\overline f_1]\mathcal O^{\mathbb H}_{\Lambda}[g_1]\big](w) \right\}\exp\left\{\frac{e_2}{2B}\left(2w(Dp-Bq)-Dw^2\right)\right\}\\\\
&=\left\{\big[\mathcal O^{\mathbb H}_{\Lambda}[f_1]\mathcal O^{\mathbb H}_{\Lambda}[g_1]- \overline{\mathcal O^{\mathbb H}_{\Lambda}[ f_2]}\mathcal O^{\mathbb H}_{\Lambda}[g_2]\big](w)\right.\\\\
&\qquad\left.e_2\big[\mathcal O^{\mathbb H}_{\Lambda}[f_2]\mathcal O^{\mathbb H}_{\Lambda}[g_1]+\overline{\mathcal O^{\mathbb H}_{\Lambda}[ f_1]}\mathcal O^{\mathbb H}_{\Lambda}[g_1]\big](w) \right\}\exp\left\{\frac{e_2}{2B}\left(2w(Dp-Bq)-Dw^2\right)\right\}\\\\
&=\mathcal O^{\mathbb H}_{\Lambda}[f](w)\mathcal O^{\mathbb H}_{\Lambda}[g](w)\exp\left\{\frac{e_2}{2B}\left(2w(Dp-Bq)-Dw^2\right)\right\}.
\end{align*}
Which completes the proof.
\end{proof}
\end{theorem}
\begin{definition}\label{def pro} For $f\in L^2(\mathbb R,\mathbb H)$ and $g\in L^1(\mathbb R,\mathbb H),$ define
\begin{equation}
(f\otimes g) =(f_2\otimes g_2 +\overline{\mathcal O^{\mathbb H}_{\Lambda^{-2}}[ f_1]}\otimes [g_1])+e_2(\overline{\mathcal O^{\mathbb H}_{\Lambda^{-2}}[ f_1]}\otimes[ g_2]-f_2\otimes g_1),
\end{equation}
where $ \otimes$ is the proposed definition of convolution.
\end{definition}
\begin{lemma}\label{lem pro}Under the assumptions of definition \ref{def pro}, we have
\begin{equation}
\left(\mathcal O^{\mathbb H}_{\Lambda}[f(t)]\otimes\mathcal O^{\mathbb H}_{\Lambda}[ g(t)]\right)(u)=\mathcal O^{\mathbb H}_{\Lambda}\left[f(t) g(t)\exp\left\{\frac{e_2}{2B}\left(2w(Dp-Bq)-Dw^2\right)\right\}\right]
\end{equation}
where $ \mathcal O^{\mathbb H}_{\Lambda}[f]\otimes\mathcal O^{\mathbb H}_{\Lambda}[ g]=\mathcal O^{\mathbb H}_{\Lambda}[fg]$
\end{lemma}

\begin{theorem}[Product theorem]\label{thm pro}Let $f,g$ be two given signal functions such that $f\in L^2(\mathbb R,\mathbb H)$ and $g\in L^1(\mathbb R,\mathbb H),$ , then  for all $w\in \mathbb R$ we have
\begin{equation}\label{eqn pro}
\mathcal O^{\mathbb H}_{\Lambda}[\overline {f}g]=\mathcal O^{\mathbb H}_{\Lambda}[f]\otimes\mathcal O^{\mathbb H}_{\Lambda}[g].
\end{equation}
\begin{proof}
By Definition \ref{def pro} and Lemma \ref{lem pro}, we have
\begin{align*}
\mathcal O^{\mathbb H}_{\Lambda}[\overline {f}g]&=\mathcal O^{\mathbb H}_{\Lambda}[(\overline {f_1}-e_2f_2)(g_1+e_2g_2)]\\
&=\mathcal O^{\mathbb H}_{\Lambda}[\overline {f_1}g_1]+\mathcal O^{\mathbb H}_{\Lambda}[ {f_2}g_2]+e_2\left(\mathcal O^{\mathbb H}_{\Lambda}[\overline {f_1}g_2]-\mathcal O^{\mathbb H}_{\Lambda}[ {f_2}g_1]\right)\\
&=\mathcal O^{\mathbb H}_{\Lambda}[\overline {f_1}]\otimes\mathcal O^{\mathbb H}_{\Lambda}[ {g_1}]+\mathcal O^{\mathbb H}_{\Lambda}[ {f_2}]\otimes\mathcal O^{\mathbb H}_{\Lambda}[ {g_2}]\\
&\qquad\qquad +e_2\left(\mathcal O^{\mathbb H}_{\Lambda}[\overline {f_1}]\otimes\mathcal O^{\mathbb H}_{\Lambda}[ {g_2}]- \mathcal O^{\mathbb H}_{\Lambda}[ {f_2}]\otimes\mathcal O^{\mathbb H}_{\Lambda}[ {g_1}]\right)\\
&=\overline{\mathcal O^{\mathbb H}_{\Lambda^{-1}}[{f_1}]}\otimes\mathcal O^{\mathbb H}_{\Lambda}[ {g_1}]+\mathcal O^{\mathbb H}_{\Lambda}[ {f_2}]\otimes\mathcal O^{\mathbb H}_{\Lambda}[ {g_2}]\\
&\qquad\qquad +e_2\left(\overline{\mathcal O^{\mathbb H}_{\Lambda^{-1}}[ {f_1}]}\otimes\mathcal O^{\mathbb H}_{\Lambda}[{g_2}]- \mathcal O^{\mathbb H}_{\Lambda}[ {f_2}]\otimes\mathcal O^{\mathbb H}_{\Lambda}[ {g_1}]\right)\\
&=\overline{\mathcal O^{\mathbb H}_{\Lambda^{-2}}\mathcal O^{\mathbb H}_{\Lambda}[{f_1}]}\otimes\mathcal O^{\mathbb H}_{\Lambda}[ {g_1}]+\mathcal O^{\mathbb H}_{\Lambda}[ {f_2}]\otimes\mathcal O^{\mathbb H}_{\Lambda}[ {g_2}]\\
&\qquad\qquad +e_2\left(\overline{\mathcal O^{\mathbb H}_{\Lambda^{-2}}\mathcal O^{\mathbb H}_{\Lambda}[ {f_1}]}\otimes\mathcal O^{\mathbb H}_{\Lambda}[{g_2}]- \mathcal O^{\mathbb H}_{\Lambda}[ {f_2}]\otimes\mathcal O^{\mathbb H}_{\Lambda}[ {g_1}]\right)\\
&=\mathcal O^{\mathbb H}_{\Lambda}[f]\otimes\mathcal O^{\mathbb H}_{\Lambda}[g].
\end{align*}
Which completes the proof.
\end{proof}
\end{theorem}
\parindent=0mm \vspace{.4in}

\section*{ Conclusion }
In this paper, we have proposed the definition of the novel integral transform known as the one-dimensional quaternion offset linear canonical transform (1D-QOLCT) which is embodiment of several well known signal processing tools. We then obtained Moyal's formula, convolution theorem and product theorem  for proposed transform. Our future work about convolution and corellation theorems for two-sided short-time offset linear canonical  transform and uncertainty principles for short-time quaternion offset linear canonical transform is in progress.

\section*{Declarations}
\begin{itemize}
\item  Availability of data and materials: The data is provided on the request to the authors.
\item Competing interests: The authors have no competing interests.
\item Funding: No funding was received for this work
\item Author's contribution: Both the authors equally contributed towards this work.
\item Acknowledgements: This work  is supported by the UGC-BSR Research Start Up Grant(No. F.30-498/2019(BSR)) provided by UGC, Govt. of India.

\end{itemize}

\end{document}